\documentclass[10pt, conference, letterpaper]{IEEEtran}
\usepackage{booktabs}
\usepackage[utf8]{inputenc}
\usepackage{cmap}
\usepackage[T1]{fontenc}
\usepackage{tikz}
\usepackage{amsmath,amssymb,amsfonts}
\usepackage[inline]{enumitem}
\usepackage[numbers]{natbib}

\usepackage{graphicx}
\usepackage{subcaption}
\usepackage{multirow}
\usepackage{xurl}

\usepackage{flushend}

\usepackage{siunitx}
\usepackage{xspace}

\newcommand{\firstly}[0]{\textit{i)}\xspace}
\newcommand{\secondly}[0]{\textit{ii)}\xspace}

\newcommand{\blfootnote}[1]{%
  \begingroup
  \renewcommand\thefootnote{}\footnote{#1}%
  \addtocounter{footnote}{-1}%
  \endgroup
}

\IEEEoverridecommandlockouts

\begin{document}

\date{}

\title{A Multi-Cloud View of Internet Background Radiation}

\author{
    \IEEEauthorblockN{
        Nils~Kempen\textsuperscript{\textsection}, 
        Ricky~K.~P.~Mok\textsuperscript{\dag}, 
        Bernhard~Degen\textsuperscript{\ddag},
        Syed~Mujtaba~Jafri\textsuperscript{\dag},
        Ralph~Holz\textsuperscript{\textsection\P}
    }
    \IEEEauthorblockA{
        \textsuperscript{\textsection}University of Münster, Münster, Germany\\
        \textsuperscript{\dag}CAIDA/UC San Diego, La Jolla, CA, USA\\
        \textsuperscript{\ddag}University of Twente, Enschede, The Netherlands\\
        \textsuperscript{\P}University of Sydney, Sydney, Australia
    }
}

\maketitle

\blfootnote{\textcopyright~IFIP, 2026. This is the author's version of the work.
It is posted here by permission of IFIP for your personal use. Not for
redistribution. The definitive version was published in the Proceedings of the
2026 IFIP Network Traffic Measurement and Analysis Conference (TMA~2026), ISBN
978-3-903176-83-6}

\begin{abstract}

    As services are increasingly centralized in public clouds, understanding the
    nature of Internet Background Radiation (IBR) hitting these particular
    environments is an important part of understanding their overall security
    posture. Classical network telescopes, long the cornerstone of IBR
    research, face hurdles here: their surface area is shrinking, and their
    well-known address ranges are easily avoided. We present a multi-cloud view
    of IBR in this paper. We deploy a passive, distributed network telescope
    with 336 IPs across five major cloud providers. We compare traffic from our
    cloud telescope with data from two classical telescopes, a large
    well-known (/9 + /10) and a small unknown /16, to analyze observational
    biases. To enable a fair comparison across very different telescope sizes,
    we tune a scan detection algorithm to determine appropriate thresholds. Our
    findings reveal that IBR in the cloud is strongly provider-dependent rather
    than geography-dependent, highlighting the necessity of multi-cloud
    deployments for comprehensive visibility of IBR in the cloud. Our cloud
    telescope also captures a distinct set of scanners, confirming that scanning
    activity is not uniform across the IPv4 space, and we confirm that small,
    distributed telescopes are ill-suited for observing random events like DDoS
    backscatter. Our work underscores that monitoring must evolve beyond
    classical telescopes to include diverse, multi-cloud vantage points to
    accurately capture IBR. 

\end{abstract}

\section{Introduction}

Classical network telescopes, also known as darknets, leverage unused routed IP
address space to passively capture incoming Internet traffic. As no active
users/services are present in the darknets, all received traffic is by
definition \emph{unsolicited} and often referred to as Internet Background
Radiation (IBR). A large body of research over the last two decades has analyzed
IBR to detect scanning traffic, backscatter from DoS attacks, and traffic
stemming from misconfigurations.

Unfortunately, the efficacy and sustainability of \emph{classical network
telescopes} are increasingly threatened by \firstly \textit{changes in
scanning strategies:} classical telescopes rely heavily on the
assumption that attackers select targets uniformly at random
\cite{mooreNetworkTelescopesTechnical2004}. This makes their well-known
address ranges susceptible to evasion by modern scanners and limits
their ability to observe targeted attacks. Furthermore, attackers shift
their focus away from sparsely populated IPv4 spaces toward dense,
high-value cloud infrastructure, which sometimes manifests as Highly
Responsive Prefixes (HRPs) \cite{sattlerPackedBrimInvestigating2023} due
to Internet consolidation \cite{isoc2019globalreport,
arkkoConsiderationsInternetConsolidation2019} and dynamic IP mapping
\cite{fayedTiesThatUnbind2021}, leading scanners to preferentially
target specific prefixes, and move away from uniformly random scanning 
\cite{izhikevichCloudWatchingUnderstanding2023,
richterScanningScannersSensing2019,
bensonLeveragingInternetBackground2015b}; \secondly \textit{exhaustion
of the IPv4 address space:} since network operators can no longer obtain
new address space from Regional Internet Registries, they are
increasingly reclaiming telescope address blocks to accommodate a
growing number of users and devices. For example, CAIDA's network
telescope has been reduced to only 63.56\% of its original size (a full
/8) due to internal demand and resale of IP ranges
\cite{mannelLessonsLearnedOperating2025a}.

Accordingly, researchers have begun to examine the use of public cloud platforms
as a vantage point to collect IBR
\cite{izhikevichCloudWatchingUnderstanding2023,
bortoluzziCloudTelescopeDistributed2023, pauleyDScopeCloudNativeInternet2023}.
However, we observe two gaps in the literature. Firstly, there is a lack of
studies that would compare classical network telescopes against a purely
\textit{passive} (listen-only) configuration of vantage points in the cloud. Secondly, it is not yet understood how IBR \textit{differs between cloud
providers}.

Our research aims to bridge these major gaps in prior studies. First, we
characterize the IBR \emph{passively} captured from virtual machines (VMs)
deployed across \emph{multiple} cloud providers, providing comprehensive
data that are also better comparable to those from classical network telescopes. Second, we
compare the data obtained from the cloud with two classical network telescopes
(UCSD-NT and SURF-NT) with complementary characteristics. SURF-NT, operated by a
research‑and‑education network in the Netherlands, has a smaller aperture 
and is not yet publicly known, reducing the likelihood that attackers will avoid
scanning its address space.

We designed and implemented CLOUD-NT, an Infrastructure-as-code multi-cloud
network telescope that supports IBR collection across five major cloud
providers. We deployed CLOUD-NT for two weeks, using 336 VMs spanning 143 cloud
regions.
Our key contributions are:
\begin{enumerate*}%
    \item a method for adjusting the detection-threshold parameters of Zeek's 
    scanner-identification algorithm to accommodate telescopes of varying sizes, 
    validated against known IP blocklists,
    \item an extensive analysis of IBR across different cloud providers,
    \item an in-depth comparison and analysis of scanners targeting cloud versus classical telescopes. 
\end{enumerate*}

We find that:
\begin{enumerate*}

    \item The IBR collected within the same cloud provider shows higher
        similarity in source IPs than IBR collected across
        different providers in the same region (average Jaccard index of \num{0.346} vs. \num{0.293}).

    \item We identify a total of \num{941}k scanners across all telescopes.
        Only \num{0.6}\% of these are commonly detected by all three network
        telescopes (\num{766}k/\num{96.6}k/\num{6.8}k of scanners are uniquely observed by
        UCSD-NT/SURF-NT/CLOUD-NT, respectively).

    \item We examine the destination ports targeted by the scanners and find
        that different telescopes observe different types of scanning traffic,
        with cloud deployments being heavily targeted on cloud-service
        associated ports and classical telescopes on ports associated with
        remote-control services.

\end{enumerate*}

\section{Related work}

The composition of IBR has shifted substantially over the past two decades. Pang
et al. \cite{pangCharacteristicsInternetBackground2004} characterized IBR at
four unused networks and found it dominated by self-propagating worms and
autorooters targeting Windows services such as CIFS, RPC and NetBIOS, with
backscatter from RSDoS attacks as a secondary component. Almost a decade later,
Dainotti et al. \cite{dainottiAnalysis0Stealth2012} documented a horizontal scan
of the entire IPv4 address space conducted by a botnet from approximately 3
million distinct IP addresses, using a coordinated and unusually covert scanning
strategy targeting \textit{SIP} server infrastructure. More recent studies have
reported a continued shift away from indiscriminate worm propagation toward
targeted, service-specific scanning campaigns and away from uniformly random
target selection \cite{izhikevichCloudWatchingUnderstanding2023,
	richterScanningScannersSensing2019, bensonLeveragingInternetBackground2015b}.

Deploying and operating large-scale network telescopes is increasingly costly
and challenging, motivating  researchers to investigate how reducing telescope
size affects threat visibility \cite{2026-degen-tsl} and explore alternative
IBR‑collection strategies. These include analyzing firewall logs from Content
Delivery Networks (CDNs) \cite{richterScanningScannersSensing2019}, analyzing
network flows traversing Internet Exchange Points
\cite{wagnerHowOperateMetaTelescope2023}, recovering IBR from ICMP errors
\cite{chanAnalyzingInternetBackground2025}, and leveraging unused IPs in active
networks (greynets) \cite{miaoExtractingInternetBackground2012}. All these
approaches require access to traffic data from production networks, which is
often unavailable publicly and may raise privacy concerns.

Cloud platforms offer researchers resources to rent (e.g., IPv4 addresses,
compute, and storage) to build network telescopes and honeypots for IBR
collection and threat intelligence. Izhikevich et al.
\cite{izhikevichCloudWatchingUnderstanding2023} deployed honeypots across
several networks, including the cloud, and combined them with \textit{Greynoise} data.
They show that scanners avoid networks without live services and differentiate
by geography. \textit{Cloud Telescope} \cite{bortoluzziCloudTelescopeDistributed2023}
used 26 AWS regions (1 IP per region) to gather IBR. Such a small deployment is
vulnerable to distortion by a few heavy hitters. \textit{DScope}
\cite{pauleyDScopeCloudNativeInternet2023} gathered unsolicited traffic on
low-cost AWS spot instances, which could be frequently terminated and
reassigned, causing IP churn. The captured traffic may include spill‑over from
former users rather than genuine malicious packets. The VMs deployed in all
these prior studies reacted to ingress traffic to different degrees. For
example, the VMs used in \textit{Cloud Telescope}
\cite{bortoluzziCloudTelescopeDistributed2023} had their SSH port changed to
65535, meaning they remained reachable from the Internet. Such interactivity can
alter scanner behavior \cite{hiesgenSpokiUnveilingNew2022a,Ferrero2025}, making
the resulting data hard or impossible to compare to that obtained by classical passive network
telescopes. Both \textit{DScope} \cite{pauleyDScopeCloudNativeInternet2023} and \textit{Cloud Telescope} \cite{bortoluzziCloudTelescopeDistributed2023} deployments used only a \textit{single}
cloud provider, limiting the generalizability of their findings across the
broader cloud ecosystem.

In this work, we carefully design our experiments to eliminate potential
confounding factors, allowing us to reveal the true visibility afforded by
building network telescopes in the cloud. Our comparative analysis uses data
collected from two classical network telescopes, allowing us to provide a comprehensive
perspective on IBR.

\section{Methodology}
\label{sec:methodology}

\begin{figure}[t]
	\centering
	\includegraphics[width=\linewidth]{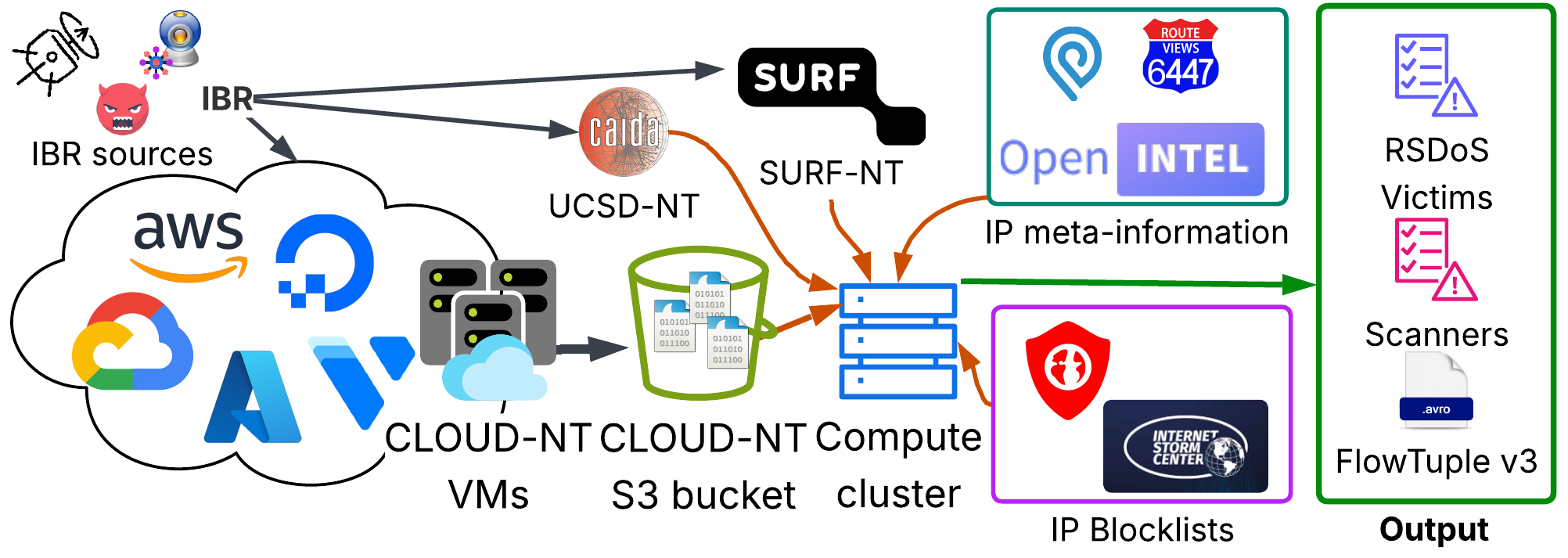}
	\caption{Our data collection and analysis pipeline integrates multiple sources of data to gather insights into IBR.}
	\label{fig:methodology_architecture} \label{fig:overview}
\end{figure}

We develop CLOUD-NT, a distributed network telescope composed of virtual
machines (VMs) across various regions of major cloud providers, to collect
IBR from April 5-23, 2025
(\S\ref{sec:method:cloudtelescope}). For
comparison, we obtain IBR data from two classical network telescopes, during
the same period (\S\ref{sec:method:darknet}). We process the data and annotate
them with additional meta-information and datasets
(\S\ref{sec:method:data_processing}). Because the telescopes differ vastly in
size, we develop methods to calibrate the thresholds for detecting scanning
events (\S\ref{sec:method:scan_detection}) and describe how we detect RSDoS
events (\S\ref{sec:method:rsdos}). Fig. \ref{fig:overview} shows an overview of
our data collection and analysis pipeline.

\subsection{Deploying a Distributed Multi-Platform Cloud Telescope}
\label{sec:method:cloudtelescope}

We design and implement CLOUD-NT, a multi-platform cloud-based network telescope
to collect IBR from diverse networks and geographic locations. Our deployment
strategy aims for provider independence, although some platform-specific
adjustments are necessary. To ensure scalability and reproducibility, we develop
a set of Terraform scripts to deploy the smallest available VMs across five
major cloud providers: AWS, GCP, Azure, Vultr, and DigitalOcean. We select these
providers to provide broad geographic coverage and diverse network topologies,
leveraging their respective strengths in region availability and pricing. To
maximize geographic and topological coverage, we deploy one VM per
availability zone to target the smallest and most granular deployment unit
available,
resulting in a total deployment of 336 instances.

Table~\ref{tab:deployment} summarizes CLOUD-NT's per-provider deployment,
including captured data volume over the measurement period and incurred costs.
The total deployment cost was approximately \num{1420.83} USD over the
measurement period $\approx$\num{78.93} USD per day.

\begin{table}[tb]
	\centering
	\caption{CLOUD-NT Deployment Overview: Per-Provider Region/Instance Counts, Capture Sizes, and Cost.}
	\label{tab:deployment}
	\setlength{\tabcolsep}{4pt}
	\begin{tabular}{l l l l r r}
		\toprule
		\textbf{Provider} & \textbf{Region} & \textbf{Instances} & \textbf{Capture} & \multicolumn{2}{c}{\textbf{Prices}\footnotemark}                                           \\
		                  &                 &                    & \textbf{Size}    & \textbf{per day}                                 & \textbf{Total}                          \\
		\midrule
		AWS               & \num{23}        & \num{73}           & \num{1.04}\,GiB  & \$\num{19.37}                                    & \$\num{348.66}                          \\
		GCP               & \num{36}        & \num{106}          & \num{1.40}\,GiB  & \$\num{25.70}                                    & \$\num{462.60}                          \\
		Azure             & \num{40}        & \num{113}          & \num{1.30}\,GiB  & \$\num{26.50}                                    & \$\num[minimum-decimal-digits = 2]{477} \\
		Vultr             & \num{30}        & \num{30}           & \num{0.62}\,GiB  & \$\num{5.365779468}                              & \$\num{96.58}                           \\
		Dig.Oc.           & \num{14}        & \num{14}           & \num{0.21}\,GiB  & \$\num[minimum-decimal-digits = 2]{1.9992}       & \$\num{35.99}                           \\
		\midrule
		\textbf{Total}    & \num{143}       & \num{336}          & \num{4.57}\,GiB  & \$\num{78.934979}                                & \$\num{1420.83}                         \\
		\bottomrule
	\end{tabular}
\end{table}

Launching many VMs in a public cloud is costly. As cloud providers
not only charge for the compute resources of the VMs but also for the use of
public IPv4 addresses, we need to be selective in where we deploy VMs. Despite the economical advantage they offer, we chose
not to use spot instances as in \cite{pauleyDScopeCloudNativeInternet2023}, as these can be terminated at any time,
which also means their IP addresses are released. In contrast to \cite{pauleyDScopeCloudNativeInternet2023}, our method allows longitudinal
observation from a stable set of IPs, so we can also distinguish between IBR
and residual (spillover) traffic caused by prior users of the IPs.

\footnotetext{Based on 2026 list prices; see Appendix \ref{app:cloud-pricing} for more details.}

We leverage IPv6 for management and data transfer to ensure the IPv4 interfaces
of the VMs are completely non-responsive to ingress traffic. Also, we configure
the cloud providers' default firewalls/security groups to allow all ingress
traffic to be forwarded to the VMs and apply \texttt{iptables} rules to prevent
the VM's OS from responding to any IPv4 traffic. We disable all VM ``health check''
services offered by the cloud providers to reduce internal probing traffic to
the VMs.

\subsection{Obtaining Data From Classical Network Telescopes}
\label{sec:method:darknet}

For comparison, we obtain the traffic data collected during the same time
period from two classical network telescopes:

\begin{enumerate*}[label=(\alph*)]
	\item \textbf{UCSD-NT:} This telescope is the largest network telescope
	      available to researchers. It is operated by CAIDA at UC San Diego,
	      capturing traffic from adjacent /9 and /10 network prefixes. Because
	      the telescope has been operational for a long time, its comprising
	      prefixes have been exposed and are widely known within the research
	      community. We use it to represent a well-established, widely used
	      vantage point in our study.

	\item \textbf{SURF-NT:} Operated by the Dutch national research and
	      education network, this telescope consists of one /16 network.
	      Because SURF-NT was set up very recently, its comprising prefixes
	      have not been exposed to the same degree. We include SURF-NT because
	      its operational setup closely resembles UCSD-NT, but its relatively
	      unexposed address space provides a valuable contrast for evaluating
	      visibility biases.
\end{enumerate*}

\subsection{Data Processing and Annotation}
\label{sec:method:data_processing}

To facilitate analysis, we convert raw packet captures collected from the cloud,
UCSD-NT, and SURF-NT into the \textit{FlowTuple} \cite{FlowTuple2021} format
using CAIDA's \textit{Corsaro} software \cite{CAIDACorsaro32024a}. This process
aggregates packets into flows at 5-minute intervals and maps source IP addresses
to Autonomous System Numbers (ASNs) using CAIDA’s prefix-to-AS mappings
\cite{RouteViewsUniversityOregon2025}.

To be able to later analyze the geographic locations and network types of IBR
sources, we annotate the flow data with geolocation and ASN type classifications
using IPinfo \cite{IPinfoTrustedIP}, which prior evaluations have shown to
provide accurate results \cite{darwichReplicationPubliclyAvailable2023}.
Additionally, we resolve the hostnames of source IPs using historical reverse
DNS (PTR) records collected by OpenINTEL \cite{OpenINTELActiveDNSa}. This allows
us to identify the specific infrastructure and hostnames associated with the
sources of IBR and extract geo-information included in PTR records using CAIDA’s
\textit{Hoiho} \cite{luckieLearningExtractGeographic2021}. We provide an overview of the
used
enrichment datasets in Table
~\ref{tab:telescope_add_data};
it additionally includes the FireHOL blocklist \cite{tsaousisFireHOLIPLists} and
the ISC SANS research scanner list \cite{centerSANSeduInternetStorm}, which we
use as proxy ground truth for the scanner-detection calibration in
\S\ref{sec:method:scan_detection}.

\subsection{Detecting Internet Scanners and RSDoS Attack Events}
\label{sec:method:scan_detection}

To detect scanning and Randomly Spoofed Denial of Service (RSDoS) attacks in
IBR, allowing for comparison between the different network telescopes, we need
an adaptable method to detect scans in different telescope setups.

To identify scanning activities, we adopt a prevalent heuristic that is derived
from the \textit{Zeek} Intrusion Detection System \cite{liReadingTeaLeaves}. Zeek's
default policy identifies a source as a scanner if it attempts to contact \num{25} or
more unique destination IP addresses on a single port within a \num{5}-minute moving
window ($N=25, T=5m$). Prior studies have adopted these parameters for
scan detection without adjusting them for the specific telescope context (e.g.,
darknets and greynets) or size (e.g., /9, /14, and /16)
\cite{liReadingTeaLeaves, mazelIdentifyingCharacterizingZMap2019}.

To maintain the rigor of threshold-based detection while ensuring accurate
temporal boundaries, we apply an overlapping sliding window workflow
\cite{lagraaDeepMiningPort2019, bou-harbCyberScanningComprehensive2014a}. Unlike
fixed-window approaches that may artificially fracture scan events across
temporal boundaries, we employ a \num{15}-minute window with \num{5}-minute increments, to
align with our \textit{FlowTuple} time intervals. This creates a \num{10}-minute
overlap between consecutive windows, guaranteeing that any isolated scan event
with a duration $D \le 10$ minutes is encapsulated within at least one
continuous processing frame, mitigating potential boundary loss effects.

We aggregate the traffic within these windows by source IPs and examine them
using a strict, top-down priority sequence. We evaluate the number of unique
destination hosts ($N_h$), ports ($N_p$), and IP-port pairs ($N_r$) contacted
against their respective thresholds ($T_h$, $T_p$, and $T_r$). Applying this
hierarchy classifies scans into three mutually exclusive categories:
\begin{enumerate*}[label=(\alph*)]
	\item \textit{Block/Random Scan} identifies high-entropy sources targeting a
	      high volume of unique destination IP and port combinations. This
	      triggers if a source exceeds the unique pairs threshold ($N_r \ge T_r$)
	      OR reaches both base thresholds in the examined window
	      ($N_h \ge T_h \land N_p \ge T_p$). This captures aggressive
	      Internet-wide sweeps that would otherwise count towards the more targeted scan metrics.
	\item \textit{Horizontal Scan} identifies sources targeting multiple hosts
	      strictly below the port threshold ($N_h \ge T_h \land N_p < T_p$).
	      This isolates horizontal scanning.
	\item \textit{Vertical Scan} identifies sources targeting a limited number of hosts
	      on many ports ($N_p \ge T_p \land N_h < T_h$). Following the
	      original Zeek implementation, we adopt $T_p = 25$ across all datasets to define this boundary.
\end{enumerate*}

This approach prevents a single massive Internet‑wide scan, one that probes many
IP‑and‑port combinations, from independently exceeding both the horizontal and
vertical thresholds, which could split the event into multiple smaller scan
records.

\subsubsection*{Threshold Calibration and Validation}

Applying the same set of threshold across all the telescope data is
inappropriate due to the vast difference in telescope sizes, ranging from \num{336}
IPs (CLOUD-NT) to over \num{10}M destination IPs (UCSD-NT).

Because modern asynchronous scanners distribute probes globally, smaller
telescopes experience a reduced statistical hit-rate per time window
compared to large subnets. For example, at a scanning rate of \num{100}k packets
per second, about \num{226.427}k and \num{7.04} packets are expected in a \num{15}-minute
window for UCSD-NT and CLOUD-NT, respectively. Thus, a static default threshold
(e.g., $T_h=25$) cannot adapt the sensitivity of the detection; it sets an
unrealistically high bar for small telescopes, causing them to miss global
scanners \cite{camargoLessMoreExploring2024}.

To set the thresholds for each telescope, we sweep a range of threshold values
for horizontal scans ($T_h \in \{2, 5, 10, 15, 25, 50, 100\}$) and random
scans ($T_r \in \{25, 50, 100, 200, 500, 1000\}$). We identify the optimal
trade-off point between scanning and other IBR with the Kneedle algorithm
to locate the point of maximum curvature (the
``elbow'') in the resulting scanner volume distributions
with per-telescope results presented in
\S\ref{sec:results:scanning}.

Furthermore, to empirically validate these derived elbows without labeled
network telescope scan data, we compare all unique source IPs extracted from the
varying threshold bounds against two IP blocklists---FireHOL blocklists
\cite{tsaousisFireHOLIPLists} and the research scanner list published by the
Internet Storm Center \cite{centerSANSeduInternetStorm}. This validation assumes
that maintaining overlap with known abuse lists indicates higher confidence in
our thresholds. Both feeds are by construction skewed toward IPs already noisy
enough to have been reported; an aggregated blocklist, like FireHOL, mitigates
individual-provider bias and establishes a practical external baseline.

\subsection{Identifying RSDoS Events and Victims}
\label{sec:method:rsdos}

Our approach to detect RSDoS attacks relies on CAIDA's \textit{Corsaro} software
and the method proposed in \cite{mooreInferringInternetDenialofservice2006a}.
Specifically, the \texttt{rsdos} plugin filters telescope traffic to isolate
unsolicited response packets, such as TCP SYN/ACKs or ICMP errors, that attack
victims inadvertently send to spoofed addresses. The pipeline then
aggregates these backscatter packets into discrete attack events by grouping
them by protocol and victim's IP addresses. To effectively eliminate
background noise and isolate genuine attacks, the system uses thresholds
regarding flow duration, packet rates, and timeout intervals. For a
comprehensive breakdown of the exact \textit{Corsaro} configuration and the specific
mathematical thresholds used in this analysis, we refer to \cite[Appendix
	J]{hiesgenAgeDDoScoveryEmpirical2024}.

\section{Results}
\label{sec:results}
We provide an overview of IBR captured by all the telescopes during our two-week
measurement period in \S\ref{sec:results:overview}. We conduct in-depth analysis
to study the similarity of IBR in different cloud providers in \S\ref{sec:results-intercloud}
and the overall cloud IBR in \S\ref{sec:results:destip}.
Next, we compare the cloud-based network telescope approach to classical network telescopes
in two different use cases: RSDoS detection in \S\ref{sec:results_rsdos},
scanner detection in \S\ref{sec:results:scanning} and scanner overlap in \S\ref{res:sec:scanner_overlap}.

\subsection{Data Overview}

Over the measurement period, UCSD-NT captured \num{51.18}\,TiB of traffic while
SURF-NT captured \num{354.2}\,GiB, compared to \num{4.57}\,GiB for CLOUD-NT
(Table~\ref{tab:deployment}).

\begin{figure}[t!]
	\centering
	\includegraphics[width=\linewidth]{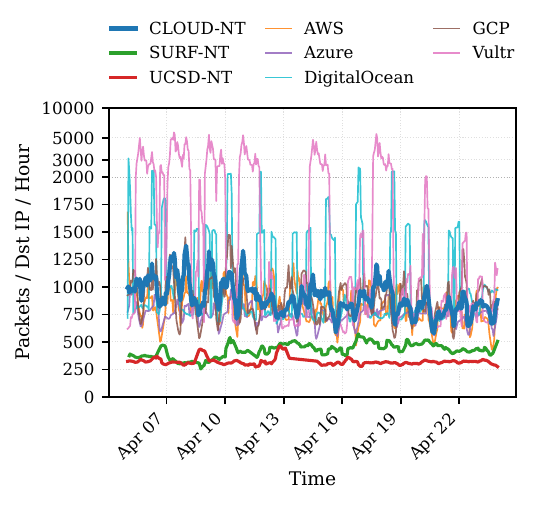}
	\caption{Traffic volume across our three network telescopes in terms of packets 
		received per hour per destination IP address. CLOUD-NT is additionally split up
		into its individual components.}
	\label{fig:traffic-timeseries}
\end{figure}

\label{sec:results:overview}
We summarize the temporal changes in volume and the source statistics of received traffic across telescopes in 
Fig.~\ref{fig:traffic-timeseries} and Table~\ref{tab:telescope_stats}, respectively.
CLOUD-NT received more unique sources per IP than the classical telescopes
(\num{46.16} vs. \num{1.14} and \num{0.09}), while the latter observed broader
source-network diversity due to their larger apertures (e.g., \num{75.34}k ASNs
for UCSD-NT). Within CLOUD-NT, the combined number of average unique source IPs
observed is lower than any single provider, since some sources target only a
subset of providers (cf. \S\ref{sec:results-intercloud}, Fig.
\ref{fig:heatmap_aggregate}). Averaging over all destinations dilutes their
contribution.
UCSD-NT and SURF-NT maintained more stable baselines than CLOUD-NT throughout
the measurement period (Fig.~\ref{fig:traffic-timeseries}), suggesting that the
variability observed in CLOUD-NT is inherent to cloud address space rather than
measurement artifacts.

\begin{table}[b!]
    \centering
    \caption{Overview of Network Telescope Datasets During the Measurement Period (April 5-23, 2025).}
    \label{tab:telescope_stats}
    \setlength{\tabcolsep}{4pt}
    \begin{tabular}{
        @{}
        l
        S[table-format=3.2]
        S[table-format=5.0, group-separator={,}, group-minimum-digits=4]
        S[table-format=5.2,  group-separator={,}, group-minimum-digits=4]
        @{}
    }
        \toprule
        {\textbf{Telescope}}
            & {\textbf{avg. uniq. src IPs /}}
            & {\textbf{uniq. ASNs}}
            & {\textbf{avg. uniq. ASNs}} \\
            & {\textbf{dst IP / h}}
            & {\textbf{overall}}
            & {\textbf{seen / h}} \\
        \midrule
        UCSD-NT            &   0.09 & 75335 & 17911.39 \\
        SURF-NT            &   1.14 & 16858 &  5059.86 \\
        CLOUD-NT (comb.)   &  46.16 & 12791 &  1545.98 \\
        \midrule
        \quad GCP          &  77.02 & 10077 &   803.67 \\
        \quad Azure        &  79.86 &  9432 &   819.94 \\
        \quad AWS          &  95.83 &  8011 &   632.20 \\
        \quad Vultr        & 156.58 &  8293 &   590.81 \\
        \quad DigitalOcean & 212.68 &  6865 &   394.80 \\
        \bottomrule
    \end{tabular}
\end{table}

Deploying VMs in cloud environments introduces unique
observational challenges. While we cannot definitively exclude the presence of
edge filtering or proprietary anti-DDoS scrubbing at the cloud provider level,
CLOUD-NT captured a sustained volume of unsolicited traffic.
Our VMs logged an average of $\approx$\num{5.7}M packets
per day, originating from $\approx$\num{16}K unique ASNs. 
We identified
$\approx$\num{5.9}M connection attempts from well-known
Internet-wide scanners, such as  Shodan, Censys, and Shadowserver, via DNS \texttt{PTR} records.
This diverse traffic profile indicates that our deployment environment was
sufficiently permissive to capture representative IBR, without noticeable filtering by the provider.  

As cloud providers routinely recycle dynamic IPv4 addresses, our VMs inevitably
captured some traffic intended for previous tenants of those IPs: ``traffic
spillover''. To quantify this, we fitted an ordinary least-squares linear
regression to each provider's daily mean packet rate over the measurement period
and tested whether the slope differed from zero using a two-sided $t$-test at
$\alpha = 0.05$. Overall, CLOUD-NT exhibited a statistically significant
downward trend of $-14.2$~packets/IP/day ($p < 0.001$, $R^2 = 0.57$), declining
from $\approx$\num{1.000}k to under~\num{800}~packets/IP. AWS showed a similarly
significant decline ($-9.3$~packets/IP/day, $p < 0.01$), while the regression
results for other providers were not statistically significant.

While we expected higher variability in Vultr and DigitalOcean due to smaller
deployments, Vultr exhibited a steep overall decline (\num{-105.6}
packets/IP/day, $p < \num{0.01}$). We identified AS~48090 (TECHOFF SRV) as the
main cause. Instead of steady traffic, AS~48090 showed a strict on-off
periodicity, alternating between inactivity and traffic bursts (peaking at
\num{41}--\num{74}\% of total packets on active days) which ceased late in our
observation window.
The cause of this on-off periodicity remains an unresolved provider-specific
artifact. We therefore interpret Vultr's regression slope as descriptive of this
measurement period rather than evidence of a stable trend.

Together, these density, stability, and per-provider-artifact differences mark
cloud IBR as qualitatively distinct from IBR in classical telescopes.

\subsection{Cloud-Provider Similarity}
\label{sec:results-intercloud}

To quantify the overlap of IBR collected across different cloud providers, we
computed the pairwise Jaccard similarity of unique source IPs across all cloud
region pairs. The resulting similarity matrix (see the full heatmap in
Appx.~\ref{sec:appendix-heatmap}) is exceptionally dense. Visualizing the
data grouped by cloud provider revealed a structural pattern. The main
diagonal blocks of the heatmap, which represent comparisons within the same
provider, showed distinctly high similarity. 
In contrast, the blocks outside this main diagonal, which represent
comparisons between different cloud providers, showed markedly lower similarity
regardless of their geographic locations.
On average,
the Jaccard similarity for intra-provider/inter-region pairs was $0.346$,
compared to $0.293$ for inter-provider/intra-region pairs. The absolute
difference of $0.053$ represents an $18\%$ relative increase in shared source
IPs within a single provider's infrastructure compared to the overlap between
different providers in the same geographic area. 

\begin{figure}[tb!]
    \centering
    \begin{subfigure}[tb]{\columnwidth}
        \includegraphics[width=.8\linewidth]{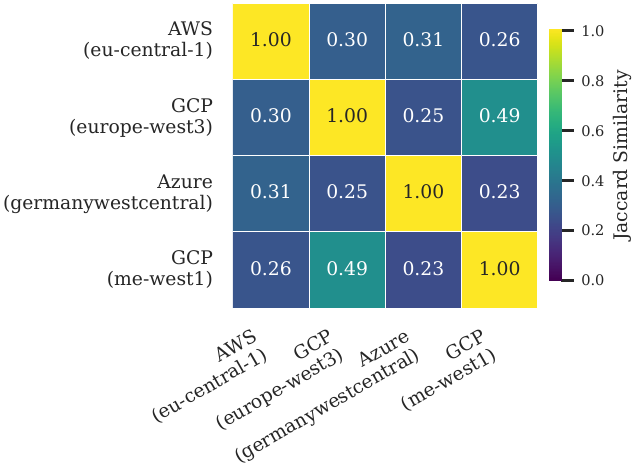}
        \caption{GCP regions and co-located competitors. High intra-provider
            similarity contrasts with low cross-provider overlap in Frankfurt.}
        \label{fig:heatmap-gcp}
    \end{subfigure}

    \vspace{1em}

    \begin{subfigure}[tb]{\columnwidth}
        \includegraphics[width=.8\linewidth]{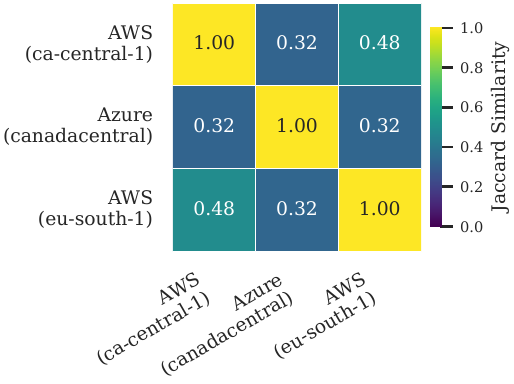}
        \caption{AWS regions and co-located competitors. Geographically distant
            AWS regions share more source IPs than co-located cross-provider
            pairs.}
        \label{fig:heatmap-aws}
    \end{subfigure}

    \caption{Jaccard similarity of observed source IPs for selected cloud
    regions. Diagonal entries (1.00) are self-comparisons. Off-diagonal
    intra-provider pairs consistently exhibit higher similarity than
    inter-provider pairs in the same geographic location.}

    \label{fig:heatmaps}
\end{figure}

As a representative example of the consistency of this phenomenon, we show the Jaccard similarities of selected region pairs in Fig.~\ref{fig:heatmaps}.

\paragraph*{Frankfurt, Germany} AWS, GCP, and Azure each have set up a region in the city. Analyzing GCP traffic in
the \texttt{europe-west3} (Frankfurt) region revealed a Jaccard similarity of
\num{0.49} with GCP's \texttt{me-west1} region (Tel Aviv) (Fig.~\ref{fig:heatmap-gcp}). In contrast, comparing
\texttt{europe-west3} to other providers physically located in the same city
yielded notably lower similarity scores: \num{0.30} with AWS \texttt{eu-central-1}
(Frankfurt) and \num{0.25} with Azure \texttt{germanywestcentral} (Frankfurt). This indicate that the GCP Frankfurt
region has nearly twice as much overlap in IBR sources with a geographically
distant GCP data center than with an Azure data center located in the same city.

\paragraph*{North America vs.\ Europe} We observed a comparable trend in AWS. The \texttt{ca-central-1} region (Canada) had a
Jaccard similarity of \num{0.48} with the geographically distant
\texttt{eu-south-1} region (Milan, Italy). In contrast, when compared to a
competitor in the same country, AWS \texttt{ca-central-1} only showed a \num{0.32}
similarity with Azure's \texttt{canadacentral} region
(Fig.~\ref{fig:heatmap-aws}). This constituted a \num{50}\% increase in
overlapping source IPs for the intra-provider pair compared to the
geographically proximate inter-provider pair.

We examined all regions where multiple providers are active and found a
consistent pattern expect for DigitalOcean's \texttt{sfo1} region.
Instead of targeting specific geographic regions across different networks, 
IBR sources primarily interacted with the
global IP space of individual cloud providers. This indicates that
an instance's IBR traffic profile depended more on its hosting
provider than on its physical geographic placement.

\subsection{Destination IP Coverage vs. Packets}
\label{sec:results:destip}

\begin{figure}[tb]
    \centering
    \includegraphics[width=.85\linewidth]{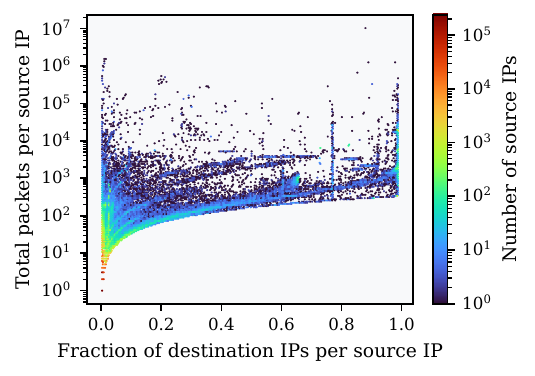}
    \caption{CLOUD-NT: heatmap of source IPs and packet counts across all 
    deployed IPs. Most sources only contact a small fraction of IPs; vertical
    artifacts arise from sources targeting a subset of all providers
    (\S\ref{sec:results:destip}).}
    \label{fig:heatmap_aggregate}
\end{figure}

Characterizing IBR captured by CLOUD-NT prior to applying any specific
behavioral thresholds, Fig. \ref{fig:heatmap_aggregate} presents a heatmap
showing the behavior of each unique source IP observed by CLOUD-NT. Adapted
from Richter and Berger
\cite{richterScanningScannersSensing2019}, source IPs are binned across two
dimensions: the fraction of monitored destination IPs contacted, or
the fan-out ($x$-axis), and the total number of packets sent ($y$-axis). The
aggregated data shows a dense concentration of source IPs clustered near the
$y$-axis. 
These hosts contacted only a small fraction of the available destination IPs but
generated a high volume of traffic, a pattern that typically indicates traffic
spillover or highly localized backscatter.
We also observed several distinct
vertical artifacts at specific destination fractions (between 0.8-1.0). 
Our investigation reveals that these vertical lines arise because some source
IPs interacted with only a subset of cloud providers. Differences in the number
of VMs deployed across providers produce these artifacts in the figure. For
example, a source scanning our entire deployed infrastructure except AWS would
contact 263 of our 336 total instances, producing the prominent vertical line
observed at $x = 0.78$. Some sources only contact a subset of providers,
matching the provider-dependency pattern from \S\ref{sec:results-intercloud}.

\subsection{RSDoS Visibility}
\label{sec:results_rsdos}

To evaluate cloud-based RSDoS monitoring, we compared the CLOUD-NT with 
SURF-NT. We found that CLOUD-NT's address space is too small for reliable
detection. SURF-NT vastly outperformed CLOUD-NT, capturing $\approx$\num{136}K
events (\num{1.33}B packets) across $\approx$\num{19.5}K unique target IPs,
compared to CLOUD-NT's $\approx$\num{15.1}K events (\num{61.1}M packets) and
\num{109} IPs. Temporally, CLOUD-NT experienced prolonged periods of zero
activity with brief spikes, in contrast to SURF-NT's continuous traffic. 
In line with \cite{2026-degen-tsl}, the visibility of RSDoS scales with the size
of the telescope and 336 IPs across five providers does not substitute for a
large classical telescope. 

\subsection{Scanner Visibility}
\label{sec:results:scanning}

\begin{figure}[tb]
    \centering
    \includegraphics[width=.6\linewidth]{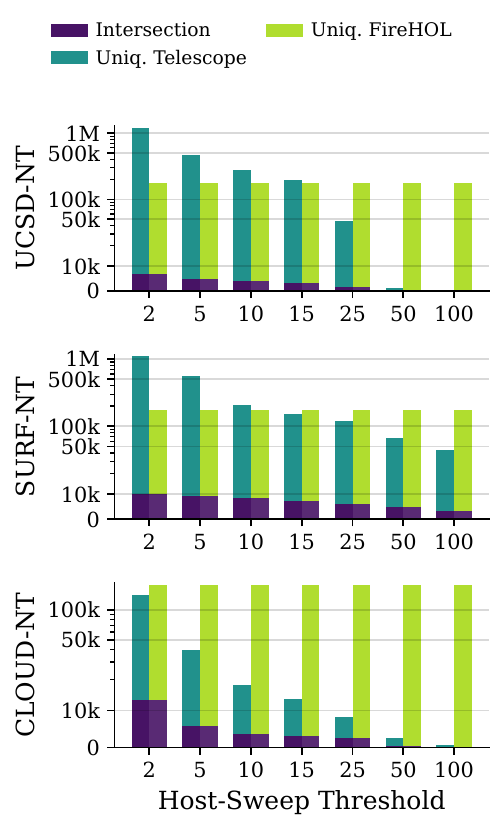}
    \caption{Intersection of identified host scanner IPs with FireHOL blocklists
        across varying host scan thresholds for all three telescopes.}
    \label{fig:overlap_analysis}
\end{figure}

As outlined in \S\ref{sec:method:scan_detection}, 
we sweep $T_h$ and $T_r$ per telescope and select the elbow point
in the retained scanner-IP curve, and cross-reference against the FireHOL 
and ISC SANS blocklists (\S\ref{sec:method:data_processing}).
Because traffic volumes varied significantly across telescopes,
we performed this calibration independently for each telescope. 
To empirically validate these selections, we cross-referenced the retained source
IPs against the \textit{FireHOL}
and \textit{ISC SANS}
blocklists (``attacks'' and ``research'' category lists, respectively). Fig. 
\ref{fig:overlap_analysis} shows the intersection and disjoint sets of IPs
across varying host scan thresholds, demonstrating how aggressive filtering
impacts the retention of known scanners versus other IBR. 
As network telescopes only capture a limited subset of Internet attacks/events,
only a fraction of the detected scanners overlap with the blocklisted IPs. 

Evaluating scanner detection across thresholds for
UCSD-NT, we observed a steep initial decline in source IPs classified
as scanners, followed by a distinct plateau. At a baseline threshold of $T_h=2$,
the telescope captured over \num{1.2}M horizontal host scanners. Increasing the
threshold to $T_h=25$ (the computed elbow) reduced this count by
96\% to \num{47.014}k. Importantly, this steep drop was not purely elimination.
While low-volume sources were discarded entirely, some IPs were simply
reclassified as port or random scanners since they no longer met the elevated
host threshold relative to their port count. Despite this overall reduction, the
intersection with the blocklists remained highly concentrated, retaining
\num{1339} IPs from FireHOL and \num{670} from ISC. Increasing the threshold
further to $T_h=50$ negatively impacted detection, dropping the total host
scanner count to \num{1098} and the FireHOL intersection to just \num{103}. For
random scans, the volume stabilized entirely at $T_r=100$. Scaling from
\num{100} up to \num{1000} resulted in less than a 1\% change in the FireHOL
overlap (\num{7716} vs.\ \num{7637}). Thus, we select thresholds of
$T_h=25$ and $T_r=100$ for UCSD-NT.

SURF-NT data demonstrated a scaled correlation between the size of
the telescope's address space and the used thresholds. Increasing the host
threshold from \num{2} to \num{15} reduced the horizontal host scanner count by
over \num{950}k IPs (dropping from $\approx$\num{1.1}M down to $\approx$\num{153}K).
This reduction in overall IP count resulted in a negligible drop in
blocklist overlap; the intersection with the ISC list only shifted from
\num{3322} at $T_h=2$ to \num{3263} at $T_h=15$ (a \num{1.7}\% loss). Because random
scans hit this mid-sized telescope more sparsely, a higher volume bound was
required to prevent targeted scans from being misclassified as random scans,
establishing calculated elbows of $T_h=15$ and $T_r=200$.

For the much smaller CLOUD-NT, employing a lower threshold of $T_h=10$ provided
the best trade-off. Compared to a baseline of \num{2}, setting $T_h=10$ reduced
the number of captured host scanners from \num{141.165}K to \num{17.593}K.
However, the FireHOL intersection remained stable at \num{3626} overlapping IPs.
If we apply UCSD-NT's threshold ($T_h=25$) to this smaller telescope, the
FireHOL overlap would drop to \num{2670}.
Therefore, we use the derived thresholds of $T_h=10$ and $T_r=100$ for CLOUD-NT.
In short, fixed default thresholds do not transfer across telescope sizes and
operators should tune them to their own telescope and check the result against
an external feed.

\subsection{Scanner Overlaps and Telescope Scale}
\label{res:sec:scanner_overlap}

\begin{figure}[tb]
    \centering
    \includegraphics[width=.95\linewidth]{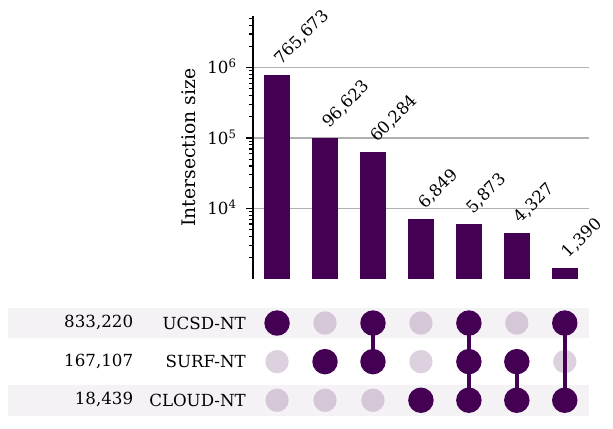}
    \caption{Scanner-IP overlap across telescopes. Only \num{5873} scanners are
    seen by all three. The telescopes are complementary, not redundant.}
    \label{fig:scanner-venn}
\end{figure}

With the chosen thresholds, we evaluated the overlap of identified scanners.
Fig.~\ref{fig:scanner-venn} shows that each telescope captured a large
exclusive set: \num{765.673}k IPs in UCSD-NT, \num{96.623}k in SURF-NT, and
\num{6.849}k in CLOUD-NT. 
Showing high marginal visibility, CLOUD-NT observed
\num{54.88} scanners per IP, compared to \num{2.4} for SURF-NT and
\num{0.066} for UCSD-NT.

Table~\ref{tab:scan_behavior} summarizes the behavior of these exclusive
scanners. While median port counts were comparable across telescopes
(\numrange{10}{32}), the tails diverged sharply. The 99\textsuperscript{th} percentile reached
\num{2438} ports in CLOUD-NT versus \num{273} in UCSD-NT and \num{37}
in SURF-NT, revealing a population of scanners that exclusively
targeted cloud prefixes with deep vertical enumeration while bypassing
classical telescopes entirely. 
We computed the Pearson correlation between ports
and hosts scanned.
We found that UCSD-NT exhibited a strong positive
correlation ($r = \num{0.68}$), indicating indiscriminate sweeps where scanners
that probe more ports also target more hosts. In contrast, CLOUD-NT showed a slight
negative correlation ($r = \num{-0.098}$, $p<0.001$), meaning scanners
probing many ports concentrate on fewer hosts, consistent with targeted
scans of specific cloud IPs.

\begin{table}[tb]
    \centering
    \caption{Exclusive Scanner Behavior by Scan Type (\S\ref{sec:method:scan_detection})}
    \begin{tabular}{llrrr}
        \toprule
        & \textbf{Type} & \textbf{Count} & \textbf{Med. Ports} & \textbf{Med. Hosts} \\
        \midrule
        \multirow{3}{*}{UCSD-NT}
            & Host   & \num{980}k  & 23 & 27 \\
            & Port   & \num{3.1}M  & 28 & 22 \\
            & Random & \num{4.3}M  & 57 & 53 \\
        \midrule
        \multirow{3}{*}{SURF-NT}
            & Host   & \num{381}k  &  4 & 31 \\
            & Port   & \num{414}   & 27 &  3 \\
            & Random & \num{95}k   & 31 & 132 \\
        \midrule
        \multirow{3}{*}{CLOUD-NT}
            & Host   & \num{30}k   &  8 & 14 \\
            & Port   & \num{1.4}k  & 32 &  9 \\
            & Random & \num{5.4}k  & 36 & 12 \\
        \bottomrule
    \end{tabular}
    \label{tab:scan_behavior}
\end{table}

\paragraph*{Targeted Ports}

To study the services targeted by scanners in cloud environments, we examined
the top destination ports of the telescopes' exclusive scanners by calculating the total number of scan events
in which each port was present (Table~\ref{tab:top_ports}). In both UCSD-NT and
SURF-NT, port 23 (Telnet) was the top port, generating nearly \num{8}M hits in
UCSD-NT alone. This predominance of Telnet traffic may be caused by IoT botnets
(e.g., Mirai) \cite{heoWhoKnockingTelnet2018}. Additionally, UCSD-NT observed
substantial traffic destined to port 8728 (MikroTik RouterOS), highlighting the
continuous automated exploitation of networking devices.
Additionally, we observed frequent scans targeting port 0. This phenomenon 
has already been extensively discussed in existing literature 
\cite{bou2014multidimensional, luchs2019curious, maghsoudlou2021zeroing} and
we did not investigate it further.

\begin{table}[tb]
    \centering
    \caption{Top 5 Targeted Ports by Telescope Exclusive Scanners (Percentage of Scan Events Including the Port)}
    \begin{tabular}{lrr|rr|rr}
        \toprule
        \textbf{Rank} & \multicolumn{2}{c|}{\textbf{UCSD-NT}} & \multicolumn{2}{c|}{\textbf{SURF-NT}} & \multicolumn{2}{c}{\textbf{CLOUD-NT}} \\
         & \textbf{Port} & \textbf{\%} & \textbf{Port} & \textbf{\%} & \textbf{Port} & \textbf{\%} \\
        \midrule
        1 & 23   & 0.60 & 23   & 3.93 & 443  & 0.04 \\
        2 & 80   & 0.46 & 80   & 2.94 & 8443 & 0.04 \\
        3 & 8728 & 0.44 & 8080 & 2.67 & 990  & 0.03 \\
        4 & 0    & 0.41 & 8000 & 2.54 & 993  & 0.03 \\
        5 & 22   & 0.39 & 8081 & 2.45 & 995  & 0.03 \\
        \bottomrule
    \end{tabular}
    \label{tab:top_ports}
\end{table}

In contrast, Telnet and HTTP were absent from the list of top ports in CLOUD-NT
(coming in on place 20 and 144 respectively).
Cloud-exclusive scanners focused on higher-level enterprise and web infrastructure. The most targeted
ports in CLOUD-NT were 443 (HTTPS) and 8443 (often used for Web Admin panels),
followed by mail servers (990, 993, 995, 110) and remote access or database
ports like 3389 (RDP) and 6379 (Redis). The presence of TLS-heavy ports implies
that cloud directed scanners would perform complete handshakes to gather
certificate data, hunt for misconfigured enterprise applications, or seek out
exposed administrative interfaces.

\paragraph*{Scanner Origins}

Table \ref{tab:top_origins} depicts the geographic and AS
origins of detected scanners. UCSD-NT traffic was highly distributed and
Asia-centric, dominated by China (35.3\%) and India (9.0\%), with the US
accounting for only 6.4\%. Furthermore, its ASN distribution lacked
concentration; each top ASNs only contributed  $\approx$0.5\% of traffic. This
lack of centralization within major hosting providers suggests the scanning
originated from widely distributed individual sources, aligning with prior
observations of decentralized IoT botnets operating heavily from residential
ISP address spaces \cite{heoWhoKnockingTelnet2018}.

\begin{table}[tb]
    \centering
    \caption{Top 3 Source Countries and ASNs of Scanners per Telescope}
    \begin{tabular}{lrl|rl}
        \toprule
        \textbf{Telescope} & \multicolumn{2}{c|}{\textbf{Top Countries (\%)}} & \multicolumn{2}{c}{\textbf{Top ASNs (\%)}} \\
        \midrule
        \multirow{3}{*}{\textbf{UCSD-NT}} 
         & CN & 35.36 & Microsoft (AS8075) & 0.53 \\
         & IN &  9.01 & Censys (AS398324)  & 0.52 \\
         & US &  6.40 & SS-Net (AS204428)  & 0.52 \\
        \midrule
        \multirow{3}{*}{\textbf{SURF-NT}} 
         & US & 20.32 & DigitalOcean (AS14061)  & 42.28 \\
         & IN & 11.76 & HiNet (AS3462)          & 10.87 \\
         & TW & 11.62 & NIB India (AS9829)      &  5.96 \\
        \midrule
        \multirow{3}{*}{\textbf{CLOUD-NT}} 
         & US & 73.34 & Linode (AS63949) & 45.58 \\
         & CN &  5.27 & Microsoft (AS8075)      & 14.31 \\
         & BR &  4.79 & Amazon (AS16509)        &  7.10 \\
        \bottomrule
    \end{tabular}
    \label{tab:top_origins}
\end{table}

We observed a cloud-on-cloud scanning phenomenon in CLOUD-NT. To ensure this
observation was not an artifact of cloud-internal traffic, we compiled IPv4 prefixes for each cloud provider and removed all intra-provider traffic
(\emph{i.e.}, traffic where the source IP belongs to the same provider hosting the
receiving telescope VM). Regardless of the specific provider evaluated, 
incoming scans were predominantly of U.S. origin ($\approx$\num{73}--\num{80}\%). The
ASN distribution was highly concentrated among competing cloud providers: ASN63949
(Linode/Akamai) alone accounted for roughly \num{45}\% of all
scanners hitting CLOUD-NT, followed closely by Microsoft (AS8075)
and AWS (AS16509). Among the top \num{1000} scanner ASes,
 \num{82.6}\% of all CLOUD-NT scanners originated from hosting ASes,
compared to only \num{39.5}\% for UCSD-NT.

The three telescopes thus see largely disjoint scanner populations. CLOUD-NT
misses the sparse, randomized scans characteristic of botnets, which are visible
to UCSD-NT, but it exposes targeted vertical scanning from commercial
datacenters aimed at enterprise and web infrastructure. CLOUD-NT demonstrates
that location in the address space can be as important as network size for
observing modern scanning campaigns.

\section{Limitations}
\label{sec:limitations}

Our study has several limitations: 
\begin{enumerate*}

    \item Our measurement only spans a bit over two weeks,
    which is insufficient to capture seasonal patterns or longer-term shifts in
    scanning behavior. Our similarity and scanner overlap analyses likewise reflect
    a single observation window. How these evolve over longer periods,
    remains future work.

    \item CLOUD-NT's aperture of 336 IPs inherently limits visibility into rare,
    randomly-distributed events such as RSDoS backscatter (\S\ref{sec:results_rsdos})
    and sparse scans. As we chose to only deploy one VM per region / availability
    zone for each provider, results for individual regions should be interpreted
    with caution, while our cross-provider findings remain robust.

    \item Our spillover analysis (in \S\ref{sec:results:overview}) reports a single
    regression slope per provider and thus captures only the aggregate decay trend.
    An address that previously hosted a popular public-facing service, with PTR
    records and client-side DNS caches still pointing to it, would plausibly receive
    residual traffic far longer than one previously used only for internal or
    outbound workloads. A per-IP characterization would refine this estimate.

    \item As discussed in \S\ref{sec:results:overview}, we cannot rule out
    provider-side DDoS scrubbing or edge filtering, but the consistent presence of
    well-known scanners in our captures provide a lower-bound for actual
    traffic. The traffic captured still accurately represents what actually is
    observed by cloud tenants.

\end{enumerate*}

\section{Discussion}
\label{sec:discussion}

Our results show that cloud and classical telescopes observe different scanning
activity. 
The scanner overlap between the telescopes is small, indicating
that neither approach subsumes the other. 
Classical telescopes capture broad
sweeps from distributed residential networks, while cloud telescopes show some
targeted vertical enumeration from commercial hosting infrastructure.
A researcher using only one type would miss the specific scans visible to the
other. For studying threats to cloud-hosted services, a cloud telescope is
necessary. Cloud-exclusive scanners probe enterprise services that are barely
visible in classical telescope traffic (Table~\ref{tab:top_ports}). The slightly
negative port-host correlation in CLOUD-NT further hints at deliberate scanning
campaigns rather than other IBR. Classical telescopes do not show this activity
because some of these scanners focus on cloud providers' address ranges.
Conversely, cloud telescopes are inadequate for RSDoS detection and for tracking
globally distributed botnets.%

The cloud provider similarity analysis has a direct operational implication.
Since IBR profiles depend more on the hosting provider than on geographic
location, deploying a cloud-based network telescope in only one cloud provider
introduces a systematic observation bias. Comprehensive cloud threat monitoring
therefore requires capturing IBR across multiple providers, not just regions.

Methodologically, we use external blocklists as a proxy ground truth to
determine scanner detection thresholds. Albeit imperfect, this empirical baseline
offers a better alternative to arbitrary parameter tuning, and demonstrates
that telescopes capture many active scanners missing from public feeds.

Cloud telescopes also offer a favorable cost-visibility tradeoff. The scanners
observed per monitored IP in CLOUD-NT exceed those in UCSD-NT by three orders of
magnitude. While scanners-per-IP is an unconventional metric for classical
telescopes, it serves as a practical measure of cost-efficiency in cloud
environments where operators pay per allocated address. This efficiency arises
because cloud IP space is actively targeted, unlike the IPv4 prefixes that host
classical telescopes. A subsampling analysis (Appendix~\ref{app:cloud-pricing})
further shows sub-linear scaling. Over our measurement period, each
additional USD of VM rental yielded $\approx$\num{1000} new source IPs at
moderate deployment sizes, falling to $\approx$\num{350} at full scale, while
\num{200} VMs already recover $\approx$\num{74}\% of full-scale visibility.

Our results suggest that a community-run cloud telescope would provide benefit
to the research community. The funding seems to be an entirely solvable issue
if several groups partner or win small-sum support of various cloud providers.
Our setup already demonstrated meaningful visibility at modest cost. We are
aware that providers generally hesitate to share data about attacks on their
networks, but releasing such data from a community telescope to the academic
community, under a framework governing use of the data, disclosure and academic
publications, seems unlikely to be a major concern.

\section{Summary}
This paper compares Internet Background Radiation (IBR) captured by a
multi-provider cloud network telescope (CLOUD-NT) with two classical network
telescopes (UCSD-NT and SURF-NT). We found that IBR is highly
provider-dependent, with scanners frequently targeting specific cloud
environments rather than geographic regions. While classical telescopes remain
necessary for observing widespread background radiation and RSDoS attacks, they
could miss targeted enterprise scans. Some scanners targeting the cloud exhibited
deep vertical enumeration, primarily originating from other hosting providers
rather than ISPs, and were often absent from IP blocklists. Our findings demonstrate that multi-cloud deployments provide a
highly efficient complement to classical darknets for monitoring
modern, service-specific threats.

\section*{Acknowledgements}

We thank our shepherd and the anonymous reviewers for their constructive
feedback. We also extend gratitude to SURF for providing the telescope data and
to the CATRIN project for providing additional data. This material is based on
research sponsored by the National Science Foundation (NSF) grants CNS-2212241,
CNS-2450552, OAC-2319959, and OAC-2531134.  Parts of the calculations for this
publication were performed on the HPC cluster PALMA II of the University of
Münster, subsidised by the DFG (INST 211/667-1). The views and conclusions
contained herein are those of the authors and should not be interpreted as
necessarily representing the official policies or endorsements, either expressed
or implied, of the funding agencies.

\bibliographystyle{IEEEtran}
\bibliography{lib-short}
\appendix

\section*{Complete Jaccard Similarity Heatmap}
\label{sec:appendix-heatmap}

\begin{figure*}[tb]
	\centering
	\includegraphics[scale=.21]{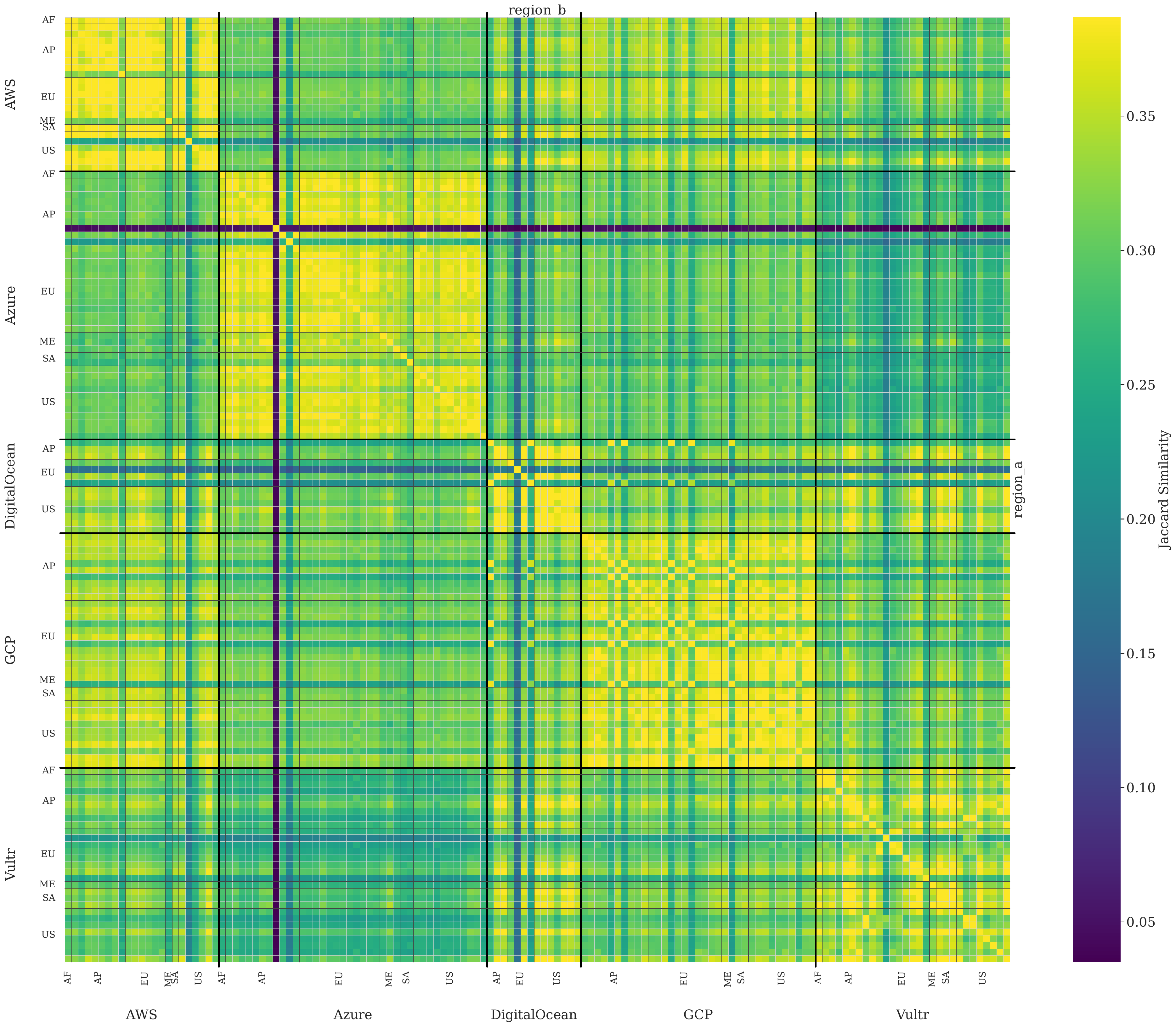}
	\caption{Complete Jaccard similarity matrix of source IPs across all cloud
		regions. Diagonal blocks (intra-provider) show markedly higher similarity
		than off-diagonal blocks, even for co-located data centers.}
	\label{fig:full-heatmap}
    \vspace{-1em}
\end{figure*}

Fig.~\ref{fig:full-heatmap} shows the complete Jaccard similarity matrix across
all monitored cloud regions. The diagonal structure indicates that
intra-provider source IP overlap consistently exceeds inter-provider overlap,
regardless of geographic proximity. Regions are grouped by provider (e.g. AWS,
Azure, GCP) to highlight this effect. We also identified outliers in the
data. The dark line in \textit{Azure} represents the \texttt{koreacentral}
region. This region is paired with the \texttt{koreasouth} region, which did
not display the same anomaly. Even though our VMs in \texttt{koreacentral} did
not share the same announced prefix as other Azure nodes, they were located
within the same AS (8075) and used IP addresses from shared prefixes. An
analysis of the source countries of the IP addresses contacting these VMs
revealed a distinct difference; the \texttt{koreacentral} region received
vastly more traffic from China than any other \textit{Azure} region. In fact,
we observed more than 30 times the number of source IPv4 addresses located in
China in \texttt{koreacentral} than we observed on average across all other
\textit{Azure} regions. Similarly, we observed outliers in
\textit{DigitalOcean's} \texttt{ams3} and \texttt{lon1} regions, where we
found substantially more source IPs originating from Pakistan, Ireland, and
Nigeria (\texttt{ams3}) or from Argentina and France (\texttt{lon1}).

\begin{table}[tb]
	\centering
	\caption{Overview of Reference Datasets}
	\label{tab:telescope_add_data}
	\begin{tabular}{lll}
		\toprule
		\textbf{Dataset Name} & \textbf{Information Extracted} & \textbf{Dataset Date} \\
		\midrule
		IPInfo                & Geolocation, ASN type          & 2025-04-28            \\
		IP2Asn                & Autonomous System Number (ASN) & 2025-04-18            \\
		OpenIntel             & Reverse DNS records (PTRs)     & 2025-04-24            \\
		Hoiho                 & Reverse DNS context (PTRs)     & -                     \\
		FireHOL               & Blocklisted IPs                & 2025-04-18            \\
		ISC SANS              & Research scanner IPs           & 2026-02-09            \\
		\bottomrule
	\end{tabular}
    \vspace{.5em}
	\caption*{ISC does not offer historical lists. We filter to only include IPs
		already present during our observation period.}
    \vspace{-1em}
\end{table}

\section*{CLOUD-NT}

\begin{figure}[ht]
	\centering
	\includegraphics[scale=.64]{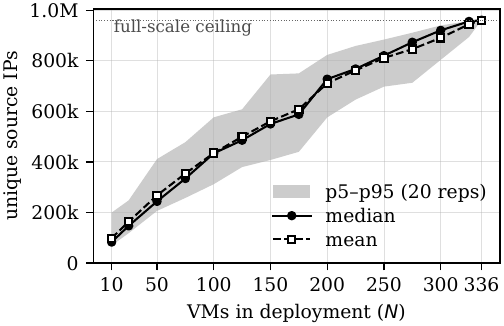}
	\caption{Unique src. IPs observed when subsampling CLOUD-NT to $N$ VMs
		(\num{20} random draws per $N$; band: p5--p95).}
	\label{fig:subsampling}
    \vspace{-1em}
\end{figure}

\vspace{-0.5em}
\paragraph*{Pricing} \label{app:cloud-pricing}
Although we deployed CLOUD-NT in 2025, we evaluate costs using 2026 regional
list prices, without any discounts or free-tiers. The GCP \textit{e2-micro}
instances (\num{2} shared vCPUs, \num{1}~GB RAM) total approximately
\num{25.70}~USD per day. Azure's equivalent \textit{Standard\_B2ats\_v2}
yields a daily cost of roughly \num{26.50}~USD, while AWS's comparable
\textit{t3a.micro} costs \num{19.37}~USD. Furthermore, our deployment
utilizing DigitalOcean's smallest droplets costs \num{1.99}~USD per day, and
the Vultr \textit{vc2-1c-1gb} instances amount to a total daily cost of
\num{5.37}~USD. Across all providers, operating the full \num{336}-VM
deployment for our measurement period would cost approximately
\num{1421}~USD in 2026.
To translate these per-VM costs into a cost-per-visibility figure, we subsampled
the CLOUD-NT VM set to $N \in \{\num{10}, \num{25}, \ldots, \num{325},
\num{336}\}$, drawing \num{20} uniform-random samples per size and counting
distinct source IPs over the full observation window. Fig.~\ref{fig:subsampling}
shows that visibility scales sub-linearly. Each USD of VM rental yields
$\approx$\num{2300} unique source IPs at small deployments, falling to
$\approx$\num{680} at full scale, and \num{200} randomly drawn VMs already
recover $\approx$\num{74}\% of full-scale visibility. Note that uniform IP-level
subsampling treats all VMs equally and therefore underestimates the value of
adding additional providers.

\paragraph*{Deployment}
The necessary \textit{Terraform} scripts to recreate our cloud deployment 
are 
available
at:
\texttt{\url{https://github.com/CAIDA/cloudtelescope-public}}.

\section*{Ethical Considerations}
In this study, we adhered to the ethical guidelines established by the Menlo Report
\cite{bailey2012menlo}. Our data collection strictly involved unsolicited
Internet Background Radiation (IBR) using well-established network telescope
methodologies prevalent over the past two decades
\cite{mooreNetworkTelescopesTechnical2004}. Because these IPs hosted no legitimate
services, the captured traffic primarily consists of automated scanning,
backscatter, and misconfigurations. In the event that misguided legitimate
traffic reached our monitors, privacy was preserved by strictly analyzing flow
data rather than inspecting packet payloads. We minimized potential harm to the
network ecosystem by ensuring our network telescopes were strictly passive;
unlike interactive honeypots, our deployments did not respond to, provoke, or
amplify malicious traffic. All deployments were conducted in compliance with the
Terms of Service and Acceptable Use Policies of the respective cloud providers,
and the resulting insights are shared to improve the understanding of IBR.

\end{document}